\documentclass[prl,aps,twocolumn]{revtex4}

\usepackage{amsbsy,amssymb,amsmath,bm}
\usepackage{epsfig,rotate}
\usepackage{fancyhdr}
\usepackage{url}
\usepackage{color}
\usepackage{graphicx}
\usepackage{dcolumn}
\usepackage{bm}
\usepackage{amssymb}
\usepackage{hyperref}

\begin{document}

\title{Memory effect and phase transition in a hierarchical trap model for spin glass}

\author{Depei Zhang}
\email{dz3vg@virginia.edu}
\affiliation{Department of Physics, University of Virginia, Charlottesville, VA 22904, USA}

\author{Tianran Chen}
\author{Marija Vucelja}
\author{Seung-Hun Lee}
\affiliation{Department of Physics, University of Virginia, Charlottesville, VA 22904, USA}

\author{Gia-Wei Chern}
\email{gchern@virginia.edu}
\affiliation{Department of Physics, University of Virginia, Charlottesville, VA 22904, USA}

\date{\today}

\begin{abstract}
We introduce an efficient dynamical tree method that enables us, for the first time, to explicitly demonstrate thermo-remanent magnetization memory effect in a hierarchical energy landscape. Our simulation nicely reproduces the nontrivial waiting-time and waiting-temperature dependences in this non-equilibrium phenomenon. We further investigate the condensation effect, in which a small set of micro-states dominates the thermodynamic behavior, in the multi-layer trap model. Importantly, a structural phase transition of the tree is shown to coincide with the onset of condensation phenomenon. Our results underscore the importance of hierarchical structure and demonstrate the intimate relation between glassy behavior and structure of barrier trees.
\end{abstract}

\maketitle

Understanding the nature of spin-glass dynamics remains a challenging task in modern statistical and condensed matter physics. Aging phenomena, a hallmark of spin glasses at low temperatures, originate from a history-dependent relaxation dynamics of these systems~\cite{vincent97,bouchaud98}. Dynamics of a glassy state is intrinsically non-stationary: the relaxation toward a ground state slows down as the system ages. One particularly intriguing dynamical behavior related to aging is the memory effect~\cite{dupuis05}, which refers to the phenomenon that the relaxation history during the cooling process seems to be kept in memory and is retrieved upon reheating. For example, if the spin-glass is held at a constant temperature $T_w$ for a period of time before further cooling, its susceptibility exhibits a remarkable dip at $T_w$ when heated back. This type of dc-memory experiment has been employed to investigate aging behavior in a wide range of glassy systems~\cite{mathieu01,mathieu10,mamiya12,samarakoon16,samarakoon17}.

Although many dynamical models have been proposed for aging phenomena with considerable success, most of them do not directly demonstrate the memory effect. A universal conclusion from various theoretical models, including the real-space droplet model~\cite{fisher88,koper88} and the energy landscape approaches~\cite{derrida80,bouchaud92}, is that the presence of memory effect suggests multiple intrinsic energy and length scales in the glassy phase~\cite{lederman91,vicent09,10Nagel,17Marinari}. Consequently, direct demonstration of memory effect with real-space Monte Carlo simulations will be very difficult due to the multiple time and length scales involved in this phenomenon~\cite{bouchaud01}. On the other hand, multiple energy/time scales can be easily encoded into dynamical models that are based on energy landscape approaches~\cite{paladin85,ogielski85,sibani89,dyre87,arous02,moretti11,jesi17}. A canonical example is the random trap model~\cite{derrida80,bouchaud92}, in which the slow dynamics below the freezing temperature $T_f$ is described by longtime activated processes consisting of jumps among different energy minima. 

In this paper, we consider the aging dynamics of a hierarchical or multilayer trap model. Through extensive Monte Carlo simulations, we demonstrate, for the first time, memory effect within the phase-space dynamics framework. Previous studies have shown a strong correlation between dynamics of glassy clusters and characteristic features of their energy landscape~\cite{de14}. Qualitative classification of barrier trees (or disconnectivity graphs) has also been conducted, and their dynamics have been investigated~\cite{wales01,despa05}. Based on the barrier-tree representation of energy landscape~\cite{wales03,becker97,wales98,garstecki99,hordijk03}, here we show that the magnitude of the memory effect depends crucially on the structure of the barrier tree. We introduce a parameter $\lambda$ to quantitatively characterize the structure of barrier trees. Essentially, this parameter controls the branching probability of the backbone tree, which consists only of saddle points. We show the existence of a dynamical phase transition at a critical $\lambda_c$ above which glassy dynamics disappears. We further associate this critical point with the condensation phenomenon that results from a competition between energy and entropy.

\begin{figure}[t]
\includegraphics[width=0.99\columnwidth]{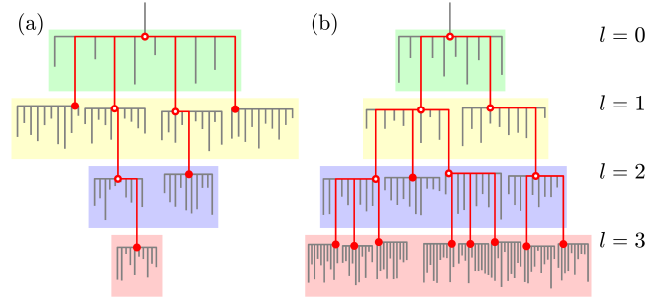}
\caption{(Color online) Schematic diagram of hierarchical barrier trees. Gray lines represent two randomly generated 4-layer barrier trees with (a) $\lambda = 0.5$ and (b) $\lambda = 2.0$. Red lines emphasize the connectivity of the all saddle points. Full (open) circles correspond to boundary (internal) saddle points.}
\label{fig:tree}
\end{figure}

We begin with a discussion of the tree representation of hierarchical energy landscape~\cite{wales03,hoffmann88}. Two examples of barrier trees are shown in Fig.~\ref{fig:tree}. Each node represents a collection of microscopic configurations with similar energies. Dynamically, we assume that micro-states lumped together into the same node can reach a local equilibrium on a short timescale compared to those it takes to access configurations belonging to other nodes~\cite{hoffmann88}. Each local minimum node (the end point of branches in Fig.~\ref{fig:tree}) in the tree represents a phase-space pocket in which the system can be trapped. The internal nodes of the tree, open and filled circles in Fig.~\ref{fig:tree}, correspond to saddle points of the landscape. The edge between two nodes indicates a possible dynamical pathway; the corresponding barrier height is indicated by the edge length. 

Following the spirit of random trap model~\cite{bouchaud92}, we assume the barrier energy $\varepsilon_l$ of local minima at $l$-th level (as well as that of saddle points at $(l+1)$-th level) is a random number drawn from an exponential distribution $\rho_l(\varepsilon_l)=e^{-\varepsilon_l/T_l }/T_l$. Importantly, this probability density gives rise to a divergent trapping time for minima at level $l$ when the temperature $T < T_l$. The characteristic temperatures $T_l$ are assumed to decrease geometrically with $l$, i.e. $T_l = T_0 \, {r}^{l}$, where $r < 1$. This provides a simple way to encode multiple energy scales into the tree structure, while maintaining a finite average total energy $\langle  E_l \rangle = \langle{\varepsilon_0}\rangle + \langle{\varepsilon_1}\rangle + \langle{\varepsilon_2} \rangle + \cdots + \langle{\varepsilon_l} \rangle$ for configurations at each level. 
In order to compute the magnetic susceptibility $\chi$, a magnetization is assigned to each node. To this end, we adopt a random magnetization model~\cite{sasaki00} and assume that the magnetization at $l$-th level is $M_l = m_0 + m_1 + m_2 + \cdots + m_l$, where $m_l$ is a random number uniformly distributed in the interval $[-\mathcal{M}_l, \mathcal{M}_l]$. Consistent with the energy barriers which become smaller with increasing level, we assume the range of magnetization $\mathcal{M}_l$ also decreases geometrically with increasing $l$.

The relaxation dynamics in a barrier tree is usually modeled by a random-walk Markov chain process. However, explicitly building a large-scale barrier tree for Monte Carlo simulations is a demanding task, which requires extremely large computer memory just for representing the tree. It is also highly inefficient as most of the nodes will not be visited by the walker. To overcome this difficulty, here we develop a {\em dynamical} tree method such that new nodes are generated {\em on the fly} according to statistical properties of the tree. 
Explicitly, the history of a random walker at level $l$ is kept in two dynamical lists: $\mathcal{L}_\varepsilon = \{ \varepsilon_0, \varepsilon_1, \varepsilon_2, \cdots, \varepsilon_l\}$ and $\mathcal{L}_m = \{m_0, m_1, m_2, \cdots, m_l\}$. These are the energy barrier and magnetization at each level, respectively. If the walker decides to make a down-transition to a lower level, random variables $\varepsilon_{l+1}$ and $m_{l+1}$ are sampled from their respective probability density and added to the respective history list. 
On the other hand, the last entries $\varepsilon_l$ and $m_l$ are deleted from the respective lists if the walker decides to go up. In doing so, we neglect the possibility that the walker will later visit exactly the same state at $l$-th level. Nonetheless, this is a reasonable approximation for barrier trees with a large number of branchings which is usually the case in the thermodynamic limit. 

Next we discuss the Markov chain process on dynamical trees. The transition probability from node $\alpha$ to $\beta$ is governed by Metropolis dynamics:
\begin{align}
	\label{eq:trans_p}
	P_{\alpha \rightarrow \beta} = Q_{\alpha \to \beta} \,\,\min\!\left\{1, e^{ - \beta\left( {\mathcal{E}}_{\beta} -  {\mathcal{E}}_{\alpha}\right)}\right\}.
\end{align}
Here $\mathcal{E}_{\alpha} = E_{\alpha} - H M_{\alpha}$ is the effective energy of node $\alpha$, $H$ is a small external probing magnetic field, $\beta \equiv 1/k_B T$ is the inverse temperature, and the probability $Q_{\alpha \to \beta}$ encodes the {\em structural} information of the tree. As in standard Metropolis dynamics, there is a finite probability $P_{\alpha \to \alpha}$ that the walker stays at the same node at every time-step; it is determined by the conservation of probability: $P_{\alpha \to \alpha} = 1 - \sum_{\beta \neq \alpha} P_{\alpha \to \beta}$.

\begin{figure}
\includegraphics[width=0.99\columnwidth]{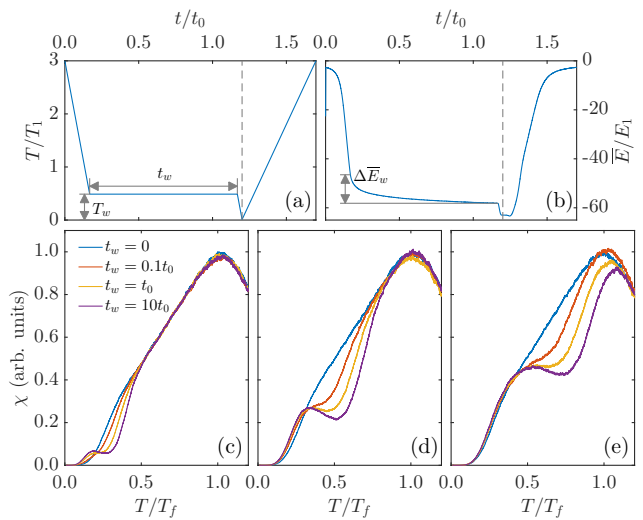}
\caption{(Color online) (a) Protocol for temperature variation for simulating memory effect in multi-layer trap model. The system is initially cooled with a constant rate until the temperature reaches $T_w$. The system then stays at $T_w$ for a finite period of waiting time $t_w$ before further cooling to a base temperature. During the subsequent measurement, a small magnetic field is applied to induce magnetization while the system is heated with a constant rate. (b) shows the system energy averaged over many independent runs as a function of time. Panels (c)--(e) show the temperature dependence of zero-field cooled DC magnetic susceptibility $\chi$,  generated by Monte Carlo simulations. The data is taken after a waiting during the cooling process with the waiting time $t_w$ in units of $5\times10^6$ MC steps at (a) $T_w = 0.2\,T_f$, (b) $T_w = 0.4\,T_f$ and (c) $T_w = 0.6\,T_f$.}
\label{fig:mc}
\end{figure}

The set of probabilities $Q_{\alpha\to\beta}$ actually defines a separate Markovian process. It can be viewed as a random walk on the tree {\em without} the energy constraint. In general, $Q_{\alpha \to \beta}$ depends on the connectivities of nodes in barrier trees. Since we do not have an explicit tree in our simulation, we assume these probabilities are given by a few parameters depending on the types of node $\alpha$ and $\beta$. Explicitly, for a walker stuck in a local minimum $\alpha$, it can only make a transition to the saddle node $\beta$ above it, or equivalently $Q_{\alpha\to \beta} = 1$. On the other hand, there are three possible transitions that can be taken at a saddle point of level $l$: the walker can go to a local minimum of the same level, or jump to a saddle node at the upper level $l-1$, or at the lower level $l+1$.  We denote the probabilities $Q_{\alpha\to\beta}$ of these three transitions as $p_0$, $p_-$, and $p_+$, respectively. 
It is worth noting that in real trees, there are unavoidably variations in these structural transition probabilities~$Q_{\alpha \to \beta}$. For example, some saddle node might have more local minima attached to it, hence a larger $p_0$. Our simplification here thus amounts to a mean-field approximation for the tree structure. 
Nonetheless, as shown below, the dynamical tree method coupled with the mean-field approximation captures the essential physics of glassy dynamics and memory phenomenon. 

We apply the above random walk simulation to study the cooling and measurement processes reported in recent experiments~\cite{samarakoon16}; also see Fig.~\ref{fig:mc}(a). The system is first cooled down from well above $T_f$ to the base temperature with a single stop at an intermediate temperature $T_w$ for some period of time $t_w$ under zero field. Once cooled down to the base temperature, the susceptibility $\chi = \langle M \rangle/H$ is measured by applying a small field upon heating at a constant rate. As demonstrated in Figs.~\ref{fig:mc}(c)-(e), our simulations successfully reproduce the memory effect which manifests itself as a prominent dip at $T_w$ when the system is heated back. Moreover, the dip becomes more pronounced with increasing waiting time $t_w$. The sensitive dependence on both waiting temperature $T_w$ and waiting time $t_{w}$ is the hallmark of memory effect in  thermo-remanent magnetization measurement~\cite{mamiya12,samarakoon16,samarakoon17}.

\begin{figure}
\includegraphics[width=0.99\columnwidth]{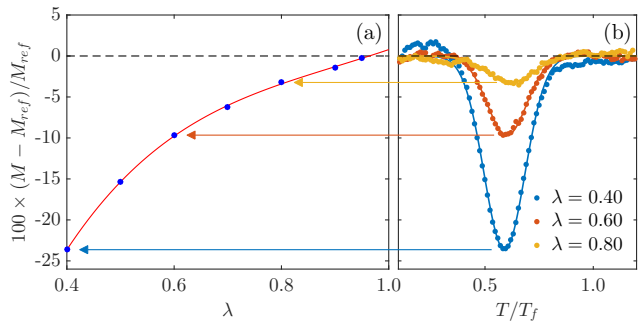}
\caption{(Color online) (a) Memory effect in terms of maximum depth of the relative change of magnetization $(M - M_{\rm ref})/M_{\rm ref}$ versus the structural parameter $\lambda = p_+/ p_-$. The red line is a guide to eye. A few examples of the temperature dependence of $(M - M_{\rm ref})/M_{\rm ref}$ are shown in panel (b).  }
\label{fig:lambda}
\end{figure}

To gain better insight of this remarkable phenomenon, we plot the average energy $\langle E \rangle$ as a function of time (in terms of Monte Carlo steps) in Fig.~\ref{fig:mc}(b). The energy $\langle E \rangle$ decreases with time initially until the cooling stops at $T_w$. During this waiting period, the walker cannot efficiently explore those levels $l'$ whose characteristic energy scales $T_{l'} \gtrsim T_w$. A longer waiting time $t_w$ at $T_w$, however, allows the walker to find energetically lower local minima in those levels $l'$, giving rise to an additional energy reduction $\Delta E_w$ from this waiting period. Upon reheating, again the average energy and level increases with time initially. As the temperature approaches $T_w$, the system needs to overcome this additional energy barrier $\Delta E_w$, leading to a dip in susceptibility.

Having demonstrated the memory effect in the hierarchical trap model, one natural question is how it is affected by the tree structure. To answer this question, we examine the dependence of memory effect on a crucial structural parameter $\lambda \equiv p_+ / p_-$, which is the average branching ratio of the backbone tree, i.e. the tree with all local minima removed (Fig.~\ref{fig:tree}). For simplicity, $p_+$, $p_-$ and hence $\lambda$ are chosen to be independent of levels $l$.  Fig.~\ref{fig:lambda}(a) shows the $\lambda$ dependence of the relative change of magnetization $(M - M_{\rm ref})/M_{\rm ref}$ that provides a quantitative measure of the memory effect, where $M$ and $M_{\rm ref}$ are the magnetization at $T_w$ with and without waiting (Fig.~\ref{fig:lambda}(b)). 
Our results show that pronounced memory effect is obtained with a small $\lambda$, corresponding to barrier trees with a lower probability of descending to lower levels. A representative example of such trees is shown in Fig.~\ref{fig:tree}(a). 

\begin{figure}
\includegraphics[width=0.99\columnwidth]{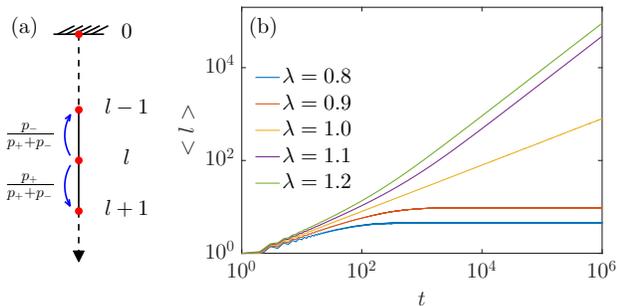}
\caption{(Color online) (a) Schematic diagram of a semi-infinite 1D random walk. In this 1D model, all the saddle nodes at the same level are treated identically, which are represented by red dots. (b) Averaged position of the walker as a function of time for different value of $\lambda = p_+ / p_-$. }
\label{fig:rw}
\end{figure}

Interestingly, the memory effect quickly disappears as $\lambda$ approaches 1, indicating a potential critical $\lambda_c$. Here we show that $\lambda_c = 1$ corresponds to a critical point of a structural transition of barrier trees. To this end, we consider a random walk process which is unaffected by the energy barrier. As discussed above, this Markovian process is governed by the transition probabilities $Q_{\alpha \to \beta}$, which only depend on the statistical property of the barrier-tree. For a walker at a local minimum $\alpha$, we have $Q_{\alpha \to \beta} = 1$, meaning that it always returns to the saddle node $\beta$ above. Consequently, we can focus only on random walks among the saddles. This effectively reduces the process to a one-dimensional (1D) random walk. At each time-step, the walker can move to a lower level with probability $\frac{p_+}{p_+ + p_-}$, or to the upper level with probability $\frac{p_-}{p_+ + p_-}$. Staying at the same level (which amounts to visiting local minima of the current level with a node-independent probability $p_0$) is ignored since it essentially changes nothing but the time unit (Fig.~\ref{fig:rw}(a)). 

A crucial observation here is that the walker cannot go above level $l=0$, which means that the random walk problem has a perfectly reflecting boundary condition at the top. Fig.~\ref{fig:rw}(b) shows the time dependence of the average position (at which level) of a walker who is initially at level-0. Here the average $\langle l \rangle$ is obtained from $10^5$ independent Monte Carlo runs. Our results clearly show two distinct dynamical regimes separated by the critical $\lambda_c = 1$. For small $\lambda < \lambda_c$, the average $\langle l \rangle$ saturates to a finite value in the large time limit. Physically, this can be understood as a balance between the tendency of the walker to move upward, and the reflection at the boundary $l = 0$. As a result of this balance, the walker never wanders too far away from the root. This is consistent with the analytical calculation showing that a walker starting at position $l \neq 0$ will always visit the root in finite time, i.e. the return probability is 1~\cite{W15}.
On the other hand, for $\lambda > \lambda_c$, the average position increases linearly with time, which is expected for a biased random walk without boundary. In the special case of $\lambda = \lambda_c$, corresponding to transition probabilities $p_+ = p_-$, we find that, even in the presence of a reflecting boundary, the walker obeys the well known time dependence $\langle l \rangle \sim t^{1/2}$ for a symmetric random walk; see Fig.~\ref{fig:rw}(b).

The above results also indicate that $\lambda_c = 1$ marks a structural phase transition for the barrier trees. First, trees with $\lambda > \lambda_c$ grow indefinitely in depth, while those with small $\lambda < \lambda_c$ tends to have finite average `length'. Although trees with large $\lambda$ have infinite number of levels, our choice of temperature parameters $\{T_l\}$ ensures that the average energy is bounded, as discussed above. The exponential distribution of barrier energy indicates that the deep-level local minima are not necessarily deep in terms of energy. Second, for random walk on the barrier tree dictated by transition probabilities $Q_{\alpha\to\beta}$, detailed balance requires that $\mathcal{N}_l \, p_+ = \mathcal{N}_{l+1} \, p_-$, where $\mathcal{N}_l$ denotes the average number of saddle nodes at level $l$. Consequently, we have $\mathcal{N}_l \sim \mathcal{N}_0 \, \lambda^l$, which means the number of saddles decreases geometrically with increasing levels for trees with $\lambda < \lambda_c = 1$, example of such trees are shown in Fig.~\ref{fig:tree}(a). Importantly, our relaxation dynamics simulations show that such sub-critical trees exhibit strong memory effect.

To further demonstrate the crucial role of tree structure on the glassy behavior, we numerically compute the average participation ratio $Y(T)$, which provides a measure of glassy behavior. It is essentially the sum of squared Boltzmann probabilities~\cite{gross84,mezard09}:
\begin{align}
    Y(T) \equiv  \left \langle \sum_\alpha W_\alpha(T)^2 \right \rangle = 
    \left \langle \frac{1}{Z^2} \sum_{\alpha} {e^{- 2 E_{\alpha}/k_BT}} \right\rangle,
\end{align}
where the summation is over all local minima, $Z = \sum_\alpha e^{- E_\alpha / k_B T}$ is the partition function, and $\langle \cdots \rangle$ denotes sample averaging. The participation ratio $Y(T)$ is used to quantitatively characterize the so-called {\em condensation} phenomenon, in which a smaller-than-exponential set of micro-states dominates the Boltzmann measure. Intuitively, the inverse $1/Y(T)$ gives an estimate of the effective number of configurations that contribute to the partition function. Consequently, when a large number of micro-states contribute equally to the Boltzmann sum, the participation ratio $Y \approx 0$. Condensation happens when the sum is dominated by a few states. 
The participation ratio can be computed analytically for the random energy model (REM)~\cite{mezard85,derrida85}, which is similar to a one-layer random trap model. In the thermodynamic limit, REM exhibits a critical temperature $T_c$, above which $Y = 0$. Condensation occurs at $T < T_c$ and the participation ratio grows linearly as $Y(T) \sim (1 - T/T_c)$.

\begin{figure}
\includegraphics[width=0.99\columnwidth]{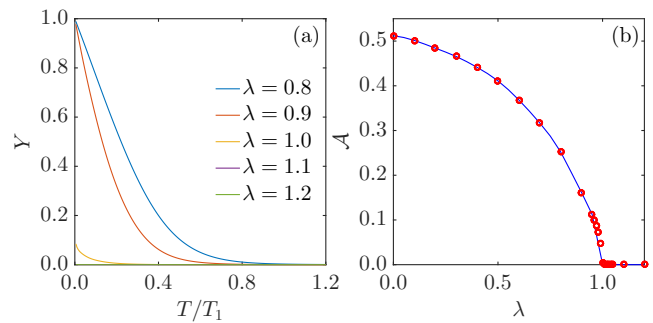}
\caption{(Color online) (a) Participation ratio $Y$ vs temperature for barrier trees obtained from the dynamical tree method. (b) A glassy order parameter $\mathcal{A}$, defined as the area under of the $Y(T)$ curve, vs the tree-structure parameter $\lambda$. }
\label{fig:pr}
\end{figure}

Here we apply our dynamical tree Monte Carlo method to numerically compute $Y(T)$ for multi-layer barrier trees. Crucially, our dynamical tree approach automatically provides thermal average to the random walk simulations. The numerical temperature dependence of $Y$ is shown in Fig.~\ref{fig:pr}(a) for varying structure factor $\lambda$. The behavior of $Y(T)$ here is similar to the REM for $\lambda < 1$. Moreover, the condensation temperature $T_c$ decreases with increasing $\lambda$. The order parameter $\mathcal{A} = \int_0^\infty Y(T) dT$, which is the area under the $Y(T)$ curve, can be used to characterize the overall degree of condensation or the glassy behavior. Consistent with our random walk simulation results, the $\lambda$ dependence of $\mathcal{A}$ (Fig.~\ref{fig:pr}(b)), indeed shows a critical point at $\lambda_c = 1$, above which the glassy order parameter vanishes. 

To summarize, we have numerically demonstrated the memory effect in a dynamical model of hierarchical barrier trees. To the best of our knowledge, this is the first numerical simulation that successfully shows the nontrivial dependence of memory effect on waiting time as well as waiting temperature. Our results strongly support the crucial role of hierarchical structure in memory effect. The hierarchical picture also indicates that the memory phenomenon is a nonequilibrium dynamics involving multiple time and length scales~\cite{lederman91,vicent09}. We further expose the crucial effect of barrier-tree structure on the memory phenomenon and the related glassy behaviors. A parameter $\lambda$ is introduced to quantitatively interpolate between different hierarchical structures such as the `palm' and `banyan' trees introduced in Ref.~\cite{becker97,wales98}. This parameter can be viewed as the branching ratio of the backbone tree that consists of only the saddle nodes.  We show that trees with a smaller branching ratio, i.e., fewer and fewer saddle points as one goes deeper, tend to exhibit a strong memory effect. 

We further establish a structural transition at the critical point $\lambda_c = 1$ above which the memory effect vanishes. In fact, trees with large branching ratio $\lambda > \lambda_c$ do not exhibit glassy behavior due to the exponential increase of the number of energetically shallow minima. This picture is supported by our result showing that condensation phenomena, in which a few deep local minima dominates the partition function, only occurs in trees with small branching ratio. The glassy transition at $\lambda_c = 1$ can also be viewed as a result of the competition between energy and entropy. While the glassy phase ($\lambda < 1$) is characterized by condensation of a few dominant micro-states, the proliferation of energetically shallow minima in trees with $\lambda > 1$ overwhelms those few deep minima, leading to the disappearance of glassy behavior and memory effect. 

\bigskip

{\em Acknowledgement}. We thank valuable discussions with P. Charbonneau, I. Klich, and L. F. Cugliandolo. S.H.L thanks the support of National Science Foundation (NSF) Grant DMR-1404994. G.-W.C. is partially supported from the Center for Materials Theory as a part of the Computational Materials Science  (CMS) program, funded by the  DOE Office of Science, Basic Energy Sciences, Materials Sciences and Engineering Division. This research was supported in part by the National Science Foundation under Grant No. NSF PHY17-48958.

\end{document}